\documentclass[conference,compsoc]{IEEEtran}
\ifCLASSOPTIONcompsoc
  \usepackage[nocompress]{cite}
\else
  \usepackage{cite}
\fi
\ifCLASSINFOpdf
  \usepackage[pdftex]{graphicx}
  \graphicspath{{../pdf/}{../jpeg/}{./figures/}}
  \DeclareGraphicsExtensions{.pdf,.jpeg,.png}
\else
\fi
\usepackage{url}

\usepackage{units}

\usepackage{etoolbox}
\newtoggle{inclIEEECopyRight}
\toggletrue{inclIEEECopyRight}

\begin{document}
\iftoggle{inclIEEECopyRight}{
    \begin{titlepage}
    \mbox{}\\{\Large \textbf{IEEE Copyright Notice}}
    \newline\newline\newline\newline
    \textcopyright~2021 IEEE. Personal use of this material is permitted.
    Permission from IEEE must be obtained for all other uses, in any current
    or future media, including reprinting/republishing this material for
    advertising or promotional purposes, creating new collective works, for
    resale or redistribution to servers or lists, or reuse of any copyrighted
    component of this work in other works.
    \newline\newline\newline\newline
    {\large Accepted to be Published in: Proceedings of the 2021 {IEEE}
    {International} {Conference} on {Cluster} {Computing} ({CLUSTER}), {EAHPC}
    {Workshop}, Sept. 7-10, 2021 Portland, Oregon, USA}
    \end{titlepage}
}{}

\bstctlcite{IEEEexample:BSTcontrol}

%
\title{A64FX -- Your Compiler You Must Decide!}

\author{\IEEEauthorblockN{Jens Domke}
\IEEEauthorblockA{RIKEN Center for Computational Science (R-CCS)\\
Kobe, Hyogo, 650-0047 Japan\\
Email: \url{http://domke.gitlab.io/##contact}}}


%


\maketitle

\begin{abstract}
The current number one of the TOP500 list, Supercomputer Fugaku, has demonstrated that CPU-only HPC
systems aren't dead and CPUs can be used for more than just being the host controller for a discrete accelerators.
While the specifications of the chip and overall system architecture, and benchmarks submitted to
various lists, like TOP500 and Green500, etc., are clearly highlighting the potential, the proliferation
of Arm into the HPC business is rather recent and hence the software stack might not be fully matured
and tuned, yet. We test 3 state-of-the-art compiler suite against a broad set of benchmarks.
Our measurements show that orders of magnitudes in performance can be gained by deviating from the
recommended usage model of the A64FX compute nodes.
\end{abstract}


%
\IEEEpeerreviewmaketitle

\section{Introduction}\label{sec:intro}
The HPC community has been testing Arm-based architectures for a few years
now~\cite{rajovic_supercomputing_2013,rajovic_mont-blanc_2016,rico_arm_2017}, and
Supercomputer Fugaku~\cite{sato_co-design_2020} is the first large-scale system in the top-end of the TOP500 list, which demonstrates the competitiveness
of Arm in
a space which recently had been dominated by Intel, AMD, and Nvidia. The benefits of Arm CPUs paired
with high bandwidth memory, as in the case of Fujitsu's A64FX processor~\cite{fujitsu_limited_fujitsu_nodate}, for the HPC field are clear:
(1) Arm CPUs are highly customizable, energy efficient, and there is an existing ecosystem of software,
compilers, tools, etc., which is readily available (unlike for the K computer with its SPARC CPU); and (2) most applications executed on HPC systems
tend to be memory-bandwidth-bound, as we have shown in a previous study~\cite{domke_double-precision_2019}.
Although, a different compute-to-bandwidth
ratio, as found in A64FX, might challenge this view in individual cases resulting in a greater influence
by the compiler onto the performance. Furthermore, despite the dominance of Arm chips in the embedded
and low-power space, the system software and compilers, such as the widely used GNU Compiler Collection
for embedded systems, might be tuned for metrics other than wide vectors (e.g., Arm's Scalable Vector
Extension) and high performance.

It is known among benchmarkers and performance analysts, that Intel’s Parallel Studio or AMD's AOCC compilers,
depending on the CPU vendor, usually yield a high baseline performance. However, for Fujitsu's A64FX, this
choice is not that obvious, yet, and options such as Fujitsu's compiler suite, GNU Compiler Collection,
Arm compilers, and HPE/Cray compilers exist, and need to be evaluated.

\begin{figure}[tbp]
    \centering
    \includegraphics[width=0.96\linewidth]{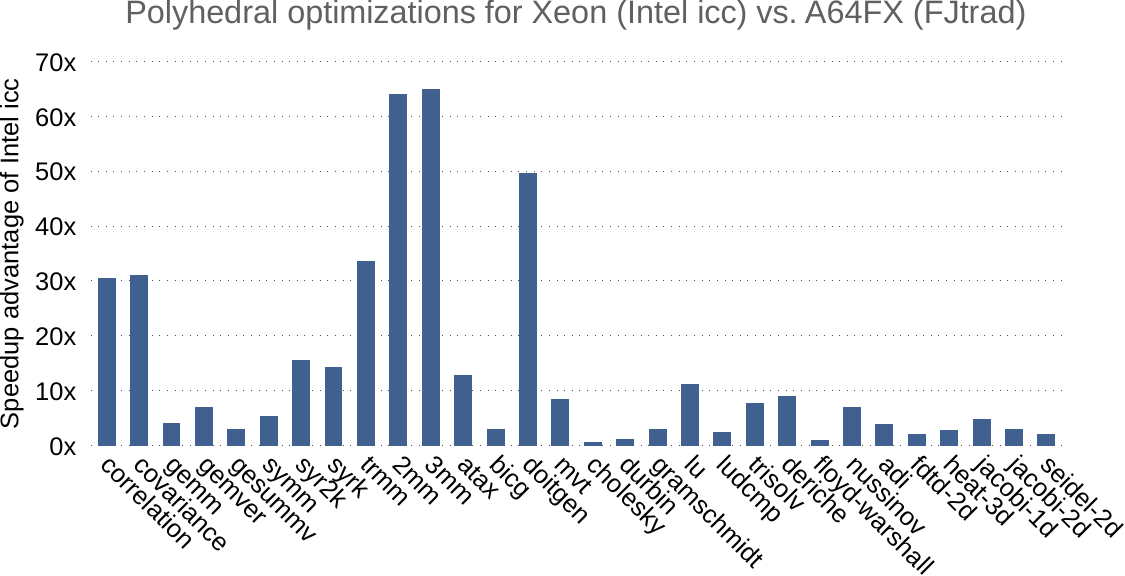}
    \caption{\textbf{Unexpected advantage of Xeon vs.~A64FX} in PolyBench[large], which \textbf{prompted this study}.
    Recommended compiler/flags used for both.}
    \label{fig:demo_polly}
\end{figure}

For example, after Supercomputer Fugaku---or short Fugaku hereafter---was put into production, we
experimented with micro kernels, namely PolyBench~\cite{pouchet_polybenchc_2016}, to debug some unexpected performance
discrepancies, especially when compared to an Intel Xeon E5-2650v4 reference CPU. While a compiler-based core-to-core
comparison for these single-threaded benchmarks is inherently inaccurate (due to ISA, caches, GHz, etc.), we did not expect
the Xeon to execute some tests up to two orders of magnitude faster, see Figure~\ref{fig:demo_polly}.
This obviously does not make sense, especially when looking at the \textit{2mm} and \textit{3mm}
matrix multiplication cases which should be compute-bound.
We will analyze the reasons in greater depth in Section~\ref{sec:eval:micro}, and it initiated
this study in which we seek to answer the following questions:
\begin{itemize}
	\item Is the recommended usage model for A64FX, i.e., compiler+flags as well as number of MPI/OMP ranks
	and threads, ideal or just a starting point?
	\item Is there a ``silver bullet'' compiler choice for A64FX?
	\item Can performance differences, compared to similar x86-based hardware, be attributed to the compiler?
\end{itemize}
To answer these questions, our contributions are as follows:
\begin{itemize}
	\item We execute a broad set of micro benchmarks, procurement proxy applications, and real-world
	application benchmarks under 5 different compiler environments to determine performance trends.
	\item A detailed discussion of the performance opportunities arising from different compiler options
	(which can also create challenges for ordinary users), and recommendations for operators and users of A64FX.
\end{itemize}

\section{Measurement Methodology}\label{sec:metho}
The performance discrepancy for the \textit{2mm} benchmark was quickly identified. Intel's C compiler
reordered the nested loop construct, while Fujitsu's C compiler (fcc) failed to do so, resulting in a
64x speedup on the Xeon core which has less than \textonehalf~of a A64FX cores's theoretical peak flop/s.
Hence, we started investigating alternative compilers to improve PolyBench and also real-world codes on Fugaku.

\subsection{Compiler and Compiler Flags}\label{sec:metho:compiler}
\textbf{FUJITSU Software Technical Computing Suite (v4.5.0)} is the recommended compiler infrastructure
for Fugaku supporting C/C++ and Fortran. It supports two modes, \textit{traditional} and \textit{clang};
the latter being based on an enhanced version of LLVM 7. We utilize both modes in this study, and link
to Fujitsu's SSL2 library for linear algebra operations whenever necessary. Most application come with
build scripts tuned for individual compilers, and we augment them with \textit{-Kfast,ocl,largepage,lto}
for performance, optimization control line (OCL) support, hugepages, and link-time-optimization, respectively.

\textbf{LLVM Compiler Infrastructure (v12)} supports C/C++ via clang; however, flang (LLVM's Fortran frontend) requires
currently a host compiler and we experienced many errors using it, and hence we skip flang and directly utilize Fujitsu's \textit{frt} compiler.
We test two settings with LLVM, the first being \textit{-Ofast -ffast-math -flto=thin} and the second
specifically targeting polyhedral optimizations with \textit{-mllvm -polly -mllvm -polly-vectorizer=polly}
and replacing the thin linker with the full linker, since \textit{thin} interfered with \textit{polly}.

\textbf{GNU Compiler Collection (v10.2.0)} supports C/C++ and Fortran, and we use
\textit{-O3 -march=native -flto} in addition to the benchmark-specific compiler flags whenever possible.

Hence, we have 5 variations, identified hereafter with \textit{FJtrad}, \textit{FJclang}, \textit{LLVM},
\textit{LLVM+Polly}, and \textit{GNU}. Unless otherwise stated, the flags listed above (or minor
variations to avoid compile/runtime issues) are in effect\footnote{Compilation scripts \&  
inputs available: \url{gitlab.com/domke/a64fxCvC}}.

Other commercial compilers from Arm (a fork of LLVM with additional optimizations and native Fortran-support)
and HPE/Cray exist; however, these are currently not available on Fugaku and we could not install
them ourselves without acquiring a license. We refer an interested reader to our related work in Section~\ref{sec:relwork}
which includes comparison for these compilers as well, but on other benchmarks/systems.

\subsection{Benchmarks -- From Micro to Macro Level}\label{sec:metho:benchmarks}
We test over 100 different kernels and scientific codes from 7 benchmark suites, as outlined hereafter:

\textbf{Micro Kernels} are a collection of 22 kernels\footnote{Source code: \url{github.com/RIKEN-RCCS/fs2020-tapp-kernels};
We use earlier snapshot; Referencing them with Kernel 1$\ldots$22 to avoid confusion.} extracted from RIKEN priority applications (see
later in this Section), which have been used during the Fugaku development for testing and validation.
These kernels are OpenMP-parallelized, primarily written in Fortran (except 5), and test various
performance-relevant aspects of one core memory group (CMG) of the A64FX processor, i.e., 12 cores (+1 assistant/OS core)
and a \unit[8]{GiB} HBM2 module.

\textbf{Polyhedral Benchmark suite} (in short, PolyBench) is a collection of 30 single-threaded
scientific kernels written in C. The input sizes can be tuned for different memory hierarchy levels,
and we use the \textit{LARGE} input (exc.: \textit{MEDIUM} for \textit{floyd-warshall}) to stress all
memory levels of A64FX.

\textbf{HPL, HPCG, and BabelStream} are commonly known to test the system's compute~\cite{dongarra_linpack_1988} and
memory performance~\cite{dongarra_hpcg_2015,deakin_gpu-stream_2016}, and are used, for example, to rank supercomputers in the TOP500 list.
HPL's and HPCG's problem sizes are configured to 36,864 and 120\textsuperscript{3}, respectively, while we
use \unit[2]{GiByte} long vectors for the stream benchmark.

\textbf{ECP proxy-apps and RIKEN Fiber mini-apps} are collections of so called \textit{proxy applications}
which are smaller representative codes and inputs for production applications commonly executed on
supercomputers in the USA and Japan. We have studied these codes previously~\cite{domke_double-precision_2019,domke_matrix_2021},
and we refer the reader to these publications for details.

\textbf{SPEC CPU[speed] and OMP} are two suites used by HPC centers and vendors to test
compute node capabilities. The former comprises 20 tests. One half are single-threaded,
integer-intensive computations and the other half tests multi-threaded, floating-point-heavy scientific
applications. The latter are 14 science workloads which are OpenMP-parallelized, too. The benchmarks
are implemented in C, Fortran, and C++, or a mix thereof. For our (non-compliant) SPEC runs, we universally
select the \textit{train}-ing input sizes.

\subsection{Evaluation Environment}\label{sec:metho:env}
All test are performed on \unit[2.2]{Ghz} A64FX-based nodes of Fugaku~\cite{fujitsu_limited_fujitsu_nodate,sato_co-design_2020}, and the benchmark's files
are cached to the first-layer storage (a SSD shared among 16 nodes) prior to its execution. We have disabled all power-saving features,
and other settings which could limit the performance of individual benchmarks. Furthermore, we submitted
all runs to the batch system with the \textit{-{}-mpi max-proc-per-node=$<$num$>$} setting, to instruct the
Fujitsu's MPI runtime to appropriately map the ranks and threads to the CMGs and cores of A64FX, i.e., \textit{spread} and \textit{close}, respectively.
The exception to this rule is PolyBench, whose tests are pinned to one core, and SPEC which comes with
its own execution environment. For SPEC, we followed a colleague's recommendation to further tune
the large page settings via \textit{XOS\_MMM\_L\_PAGING\_POLICY=demand:demand:demand} and \textit{XOS\_MMM\_L\_ARENA\_LOCK\_TYPE=0},
and specify \textit{OMP\_PROC\_BIND=close}, but left these settings at default values for all other benchmarks.

While theoretically possible, we note that tuning all our 100+ benchmarks individually with the full range of compiler flags,
runtime parameters, and manual code refactoring, etc., for optimal performance is outside the scope of this work,
which seeks to identify or disproof the existence of a ``silver bullet'' compiler for the A64FX processor.

\subsection{Measurement Approach and Metrics}\label{sec:metho:measure}
While Fujitsu's and RIKEN's recommendation for Fugaku and A64FX is 4 ranks (one per CMG) and 12 OpenMP threads per rank per node, this
may not always be ideal, since some codes prefer or even require a power-of-2 for ranks/threads (e.g.,
\textit{SWFFT}) or do not scale with number of threads (e.g., SPEC \textit{imagick}'s sweet
spot is 8 threads). Hence, we employ a exploration phase for each compiler and test various
MPI and/or OMP combinations for all parallelized, strong-scaling benchmarks
(except: weak-scaling \textit{MiniAMR} and \textit{XSBench}), using 3 trial runs each.
The fastest time-to-solution determines the final MPI/OMP setting (individual per compiler) for the
performance runs, which we run again 10 times.
We manually instrumented all benchmarks to only report the time-to-solution of the region
of interest, i.e., excluding any pre-/post-processing phases; except for SPEC CPU/OMP where
we rely on SPEC's runtime reporting. Performing 10 runs should suffice, since we experience
low run-to-run variability on A64FX. For example, AMG's coefficient of variation (CV) in runtime was
below 0.114\%, and we only see high variability in BabelStream with a CV of up to 22\% which is still
noticeably smaller than the gap between compilers.

\section{Evaluation and Result Discussion}\label{sec:eval}
\begin{figure}[tbhp]
    \centering
    \includegraphics[width=1.085\linewidth]{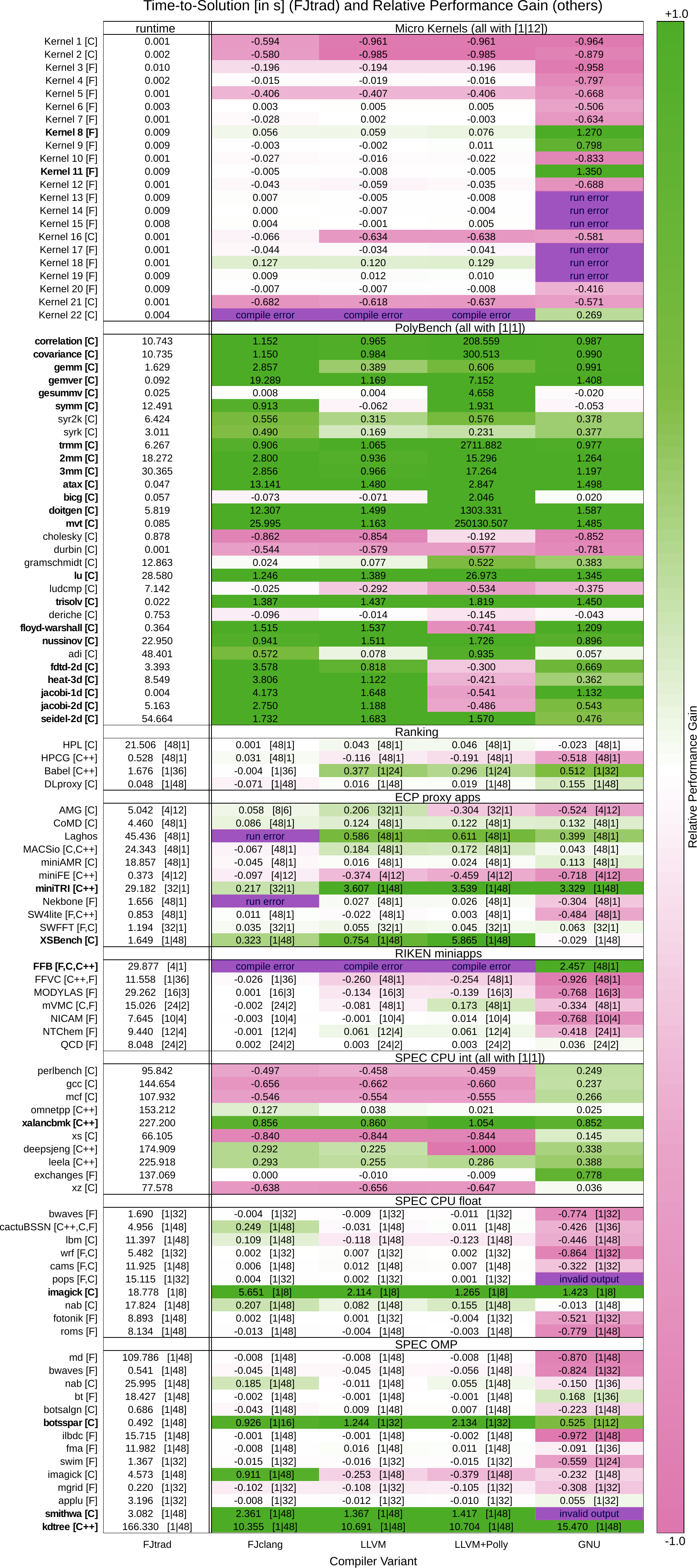}
    \caption{Absolute time-to-solution (numbers; \textit{FJtrad} column) and relative gain over \textit{FJtrad} (numbers; color coding; non-\textit{FJtrad} columns)
    for all benchmarks; Invalid entries explain (e.g. ``compiler error'', see Kernel 22); Programming
    language indicated after benchmark name; Benchmark names with $>$2x speedup in bold; Number of MPI rank \& OMP threads in brackets}
    \label{fig:compare_compiler}
\end{figure}

We report the fastest runtime across 10 \textit{performance runs}
(cf.~Sec.~\ref{sec:metho:measure}), and relative comparisons to the recommended compiler, i.e.,
using Fujitsu's compiler suite in \textit{trad} mode. Figure~\ref{fig:compare_compiler} is
additionally color-coded with the relative performance gain (see~\cite{hoefler_scientific_2015}) of each
compiler over the \textit{FJtrad} baseline, with white for similar runtime and dark green
indicating 2x speedup.  Benchmarks exhibiting over 2x speedup are further signalized with a \textbf{bold} name.
Furthermore, unsolvable compilation errors, runtime errors, and invalid runs are encoded in dark pink.
Additionally, we added the best parallelization configuration, i.e., shown via [\#MPI ranks $|$ \#OMP threads]
for each compiler/benchmark combination, except for the micro, PolyBench, and SPEC CPUint for which it is
universal and written in the header.

\subsection{Micro Kernels and Polybench}\label{sec:eval:micro}
For the Micro Kernels representing RIKEN's priority apps, we clearly see the results
of the co-design efforts for Fugaku. Fujitsu's compiler in traditional mode outperforms
all other compilers in nearly all test. Only the GNU compiler is able to noticeably beat \textit{FJtrad}
in 4 of the 22 tests, but also produces 6 executables which result in runtime errors.
Assuming that we switch always to the best compiler option, then we could reduce the runtime by 17\% on
average, with a median of 0\%, and peak of 2.4x improvement.
However, for PolyBench the roles reverse, with \textit{LLVM+Polly} showing the best results,
followed by \textit{FJclang} in some cases. Especially for \textit{mvt} the polyhedral optimizations
resulted in over 250.000x speedup. Choosing the best compiler over \textit{FJtrad} results in
a median speedup of 3.8x. It remains to be seen if \textit{polly}'s gains translate to
``real'' scientific codes hereafter.

\subsection{TOP500, ECP, and Fiber Proxys}\label{sec:eval:proxy}
HPL gains a minor advantage ($\approx$5\%) by using LLVM over Fujitsu's compilers, despite that
most of the calculations are performed within SSL2, which holds true for the convolution kernel
of the deep learning proxy, too. BabelStream shows the largest gain from switching to LLVM or GNU
with up to 51\% lower runtime. With a few exceptions, like FFB and mVMC, Fujitsu dominates the
other compilers on Fiber mini-apps, which is consistent with the Micro Kernel results shown earlier.
For ECP proxy-apps the conclusions reverses, and the user would be advised to switch to LLVM or
GNU in almost all cases. Such a change leads to an average speedup of 1.65x (median 1.09x).
The 6.7x speedup for XSBench is salient, because it also demonstrates that \textit{polly}
can have an impact on real workloads.

\subsection{SPEC CPU and OMP}\label{sec:eval:spec}
The SPEC measurements reveal multiple interesting insights. Firstly, \textit{FJtrad} outperforms
any Clang-based alternative on A64FX on integer-intensive codes; however, the GNU compiler almost
universally beats \textit{FJtrad} at the same single-threaded workloads. We speculate that this
advantage is partially a result of GNU's prevalence in the embedded space where many of the Arm CPUs have no floating-point
units, and also a result of Arm's continued investments into the open-source GNU compilers~\cite{christina_blog_2020}.
By contrast, for multi-threaded and floating-point-based SPEC CPU workloads, as well as SPEC OMP,
the GNU compiler is currently the worst choice on A64FX. Furthermore, many of the applications
are written in Fortran, and hence there is little benefit (apart from maybe link-time-optimizations)
for switching to LLVM. For C/C++ applications on the other hand, LLVM-based compilers (incl.~\textit{FJclang}),
and \textit{GNU} in a few cases, can yield a runtime benefit over \textit{FJtrad}. We see speedup as high as
16.5x in SPEC OMP simply by switching compilers (e.g., for kdtree), with an average improvement of 49\% in SPEC
CPU and 2.5x speedup in SPEC OMP. The median runtime improvement from choosing the best compiler across
both SPEC suites is 14\%.

Overall, across all 108 benchmarks and realistic workloads, we see that a median runtime improvement of 16\% is possible by
selecting an appropriate compiler, without any changes to the source code or other tuning methods.

\section{Related Work}\label{sec:relwork}
Various publications and reports of A64FX evaluations have been released recently. For
example,~\cite{huber_case_2021} tested LLVM and its SVE code generation capabilities for DCA++,
and~\cite{burford_ookami_2021} compared a limited set of applications with LLVM, GNU, ARM, and Cray
compilers and focused on SVE and multi-node scaling. Similarly,~\cite{michalowicz_comparing_2021}
investigated OpenMP-scaling of 3 proxy apps on A64FX while comparing 5 compilers,
and~\cite{poenaru_evaluation_2021} measured nearly a dozen proxy apps (different from ours) on ARM
and x86 for multiple compilers, but lacked LLVM. In~\cite{alappat_performance_2020}, the authors analyzed
the achievable bandwidth for a set of memory-bandwidth-bound kernels using GNU, and~\cite{sreepathi_e3sm_2021}
reported a 2x performance advantage of GNU compilers for the E3SM climate code. 
Lastly, the studies~\cite{jackson_investigating_2020,odajima_preliminary_2020,sato_co-design_2020}
compare A64FX with Fujitsu's traditional mode to ARM-based ThunderX2 and Intel Xeon CPUs using
various priority apps and proxies.
All these studies are complementary to our comprehensive compiler comparison for a wide
variety of workloads, since they use different apps, compilers, or evaluation approaches.

\section{Conclusion}\label{sec:concl}
In conclusion, we demonstrate a clear benefit from exploring alternative compilers for the A64FX CPU as a valid first-order
tuning method before investing a considerable amount of effort into testing ``exotic'' compiler flags, environment variables, and
performing manual code refactoring.
Especially, the performance discrepancy for PolyBench, which we show in Figure~\ref{fig:demo_polly}, was
solved by switching from the recommended \textit{FJtrad} to LLVM 12 compiler, but the \textit{polly}
optimizations seem rarely applicable or beneficial outside this benchmark set.
To revisit our initial question,
we could not identify a ``silver bullet'' compiler for A64FX, but our measurements give indications
for which compilers work well in which situations, i.e., Fujitsu for Fortran codes, GNU for integer-intensive
apps, and any clang-based compilers for C/C++. Furthermore, we noticed that for ``legacy'' applications, the
recommended usage model of 4 ranks and 12 threads per A64FX node results in suboptimal time-to-solution
more often than not, and that the Arm software ecosystem for HPC is not as mature as for x86, yet.

Our work is just one among many similar, early explorations of the newly introduced Arm-base CPU
for high-performance computing, but it gives reason to believe that potential performance deficiencies, when
directly compared against x86 for the same applications, are most likely the results of immature compilers.
Hence, our recommendation to administrators and users of A64FX-based supercomputers is to install and test as
many different, available compilers as possible to extract the true performance potential from the
A64FX CPU. Similarly, it could be worthwhile to revisit how various system libraries, such as for searching, sorting, routing,
or MPI libraries, etc., and other OS packages are (pre-)compiled for the A64FX processor.



\ifCLASSOPTIONcompsoc
  \section*{Acknowledgments}
\else
  \section*{Acknowledgment}
\fi

The authors would like to thank various colleagues from RIKEN R-CCS, in particular members of the NG-HPA, HPAI, and 
Operations team, for their feedback, discussions, and assistance with software installation and application debugging.
This work was supported by the Japan Society for the Promotion of Science KAKENHI Grant Number JP19H04119;
and by the New Energy and Industrial Technology Development Organization (NEDO).



\bibliographystyle{IEEEtran}
\bibliography{IEEEabrv,a64fxCvC}

\begin{thebibliography}{10}
\providecommand{\url}[1]{#1}
\csname url@samestyle\endcsname
\providecommand{\newblock}{\relax}
\providecommand{\bibinfo}[2]{#2}
\providecommand{\BIBentrySTDinterwordspacing}{\spaceskip=0pt\relax}
\providecommand{\BIBentryALTinterwordstretchfactor}{4}
\providecommand{\BIBentryALTinterwordspacing}{\spaceskip=\fontdimen2\font plus
\BIBentryALTinterwordstretchfactor\fontdimen3\font minus
  \fontdimen4\font\relax}
\providecommand{\BIBforeignlanguage}[2]{{%
\expandafter\ifx\csname l@#1\endcsname\relax
\typeout{** WARNING: IEEEtran.bst: No hyphenation pattern has been}%
\typeout{** loaded for the language `#1'. Using the pattern for}%
\typeout{** the default language instead.}%
\else
\language=\csname l@#1\endcsname
\fi
#2}}
\providecommand{\BIBdecl}{\relax}
\BIBdecl

\bibitem{rajovic_supercomputing_2013}
N.~Rajovic, P.~M. Carpenter, I.~Gelado, N.~Puzovic, A.~Ramirez, and M.~Valero,
  ``Supercomputing with {Commodity} {CPUs}: {Are} {Mobile} {SoCs} {Ready} for
  {HPC}?'' in \emph{Proceedings of the {International} {Conference} on {High}
  {Performance} {Computing}, {Networking}, {Storage} and {Analysis}}, ser. {SC}
  '13.\hskip 1em plus 0.5em minus 0.4em\relax New York, NY, USA: Association
  for Computing Machinery, 2013, event-place: Denver, Colorado.

\bibitem{rajovic_mont-blanc_2016}
N.~Rajovic, A.~Rico, F.~Mantovani, D.~Ruiz, J.~O. Vilarrubi, C.~Gomez,
  L.~Backes, D.~Nieto, H.~Servat, X.~Martorell, J.~Labarta, E.~Ayguade,
  C.~Adeniyi-Jones, S.~Derradji, H.~Gloaguen, P.~Lanucara, N.~Sanna, J.-F.
  Mehaut, K.~Pouget, B.~Videau, E.~Boyer, M.~Allalen, A.~Auweter, D.~Brayford,
  D.~Tafani, V.~Weinberg, D.~Br{\"o}mmel, R.~Halver, J.~H. Meinke, R.~Beivide,
  M.~Benito, E.~Vallejo, M.~Valero, and A.~Ramirez, ``The {Mont}-{Blanc}
  {Prototype}: {An} {Alternative} {Approach} for {HPC} {Systems},'' in
  \emph{{SC} '16: {Proceedings} of the {International} {Conference} for {High}
  {Performance} {Computing}, {Networking}, {Storage} and {Analysis}}, 2016, pp.
  444--455.

\bibitem{rico_arm_2017}
A.~Rico, J.~A. Joao, C.~Adeniyi-Jones, and E.~Van~Hensbergen, ``{ARM} {HPC}
  {Ecosystem} and the {Reemergence} of {Vectors}: {Invited} {Paper},'' in
  \emph{Proceedings of the {Computing} {Frontiers} {Conference}}, ser.
  {CF}'17.\hskip 1em plus 0.5em minus 0.4em\relax New York, NY, USA:
  Association for Computing Machinery, 2017, pp. 329--334.

\bibitem{sato_co-design_2020}
M.~Sato, Y.~Ishikawa, H.~Tomita, Y.~Kodama, T.~Odajima, M.~Tsuji, H.~Yashiro,
  M.~Aoki, N.~Shida, I.~Miyoshi, K.~Hirai, A.~Furuya, A.~Asato, K.~Morita, and
  T.~Shimizu, ``Co-{Design} for {A64FX} {Manycore} {Processor} and
  "{Fugaku}",'' in \emph{Proceedings of the {International} {Conference} for
  {High} {Performance} {Computing}, {Networking}, {Storage} and {Analysis}},
  ser. {SC} '20.\hskip 1em plus 0.5em minus 0.4em\relax IEEE Press, 2020,
  event-place: Atlanta, Georgia.

\bibitem{fujitsu_limited_fujitsu_nodate}
\BIBentryALTinterwordspacing
{Fujitsu Limited}, ``{FUJITSU} {Processor} {A64FX}.'' [Online]. Available:
  \url{https://www.fujitsu.com/downloads/SUPER/a64fx/a64fx_datasheet_en.pdf}
\BIBentrySTDinterwordspacing

\bibitem{domke_double-precision_2019}
J.~Domke, K.~Matsumura, M.~Wahib, H.~Zhang, K.~Yashima, T.~Tsuchikawa,
  Y.~Tsuji, A.~Podobas, and S.~Matsuoka, ``Double-precision {FPUs} in
  {High}-{Performance} {Computing}: an {Embarrassment} of {Riches}?'' in
  \emph{2019 {IEEE} {International} {Parallel} and {Distributed} {Processing}
  {Symposium}, {IPDPS} 2019, {Rio} de {Janeiro}, {Brazil}, {May} 20-24, 2019},
  Rio de Janeiro, Brazil, May 2019.

\bibitem{pouchet_polybenchc_2016}
\BIBentryALTinterwordspacing
L.-N. Pouchet and M.~Taylor, ``{PolyBench}/{C} 4.2.1 (beta),'' May 2016.
  [Online]. Available: \url{https://sourceforge.net/projects/polybench/}
\BIBentrySTDinterwordspacing

\bibitem{dongarra_linpack_1988}
\BIBentryALTinterwordspacing
J.~Dongarra, ``The {LINPACK} {Benchmark}: {An} {Explanation},'' in
  \emph{Proceedings of the 1st {International} {Conference} on
  {Supercomputing}}.\hskip 1em plus 0.5em minus 0.4em\relax London, UK, UK:
  Springer-Verlag, 1988, pp. 456--474. [Online]. Available:
  \url{http://dl.acm.org/citation.cfm?id=647970.742568}
\BIBentrySTDinterwordspacing

\bibitem{dongarra_hpcg_2015}
\BIBentryALTinterwordspacing
J.~Dongarra, M.~Heroux, and P.~Luszczek, ``{HPCG} {Benchmark}: a {New} {Metric}
  for {Ranking} {High} {Performance} {Computing} {Systems},'' University of
  Tennessee, Tech. Rep. ut-eecs-15-736, Jan. 2015. [Online]. Available:
  \url{https://library.eecs.utk.edu/pub/594}
\BIBentrySTDinterwordspacing

\bibitem{deakin_gpu-stream_2016}
T.~Deakin, J.~Price, M.~Martineau, and S.~McIntosh-Smith, ``{GPU}-{STREAM}
  v2.0: {Benchmarking} the {Achievable} {Memory} {Bandwidth} of {Many}-{Core}
  {Processors} {Across} {Diverse} {Parallel} {Programming} {Models},'' in
  \emph{High {Performance} {Computing}}, M.~Taufer, B.~Mohr, and J.~M. Kunkel,
  Eds.\hskip 1em plus 0.5em minus 0.4em\relax Cham: Springer International
  Publishing, 2016, pp. 489--507.

\bibitem{domke_matrix_2021}
J.~Domke, E.~Vatai, A.~Drozd, C.~Peng, Y.~Oyama, L.~Zhang, S.~Salaria,
  D.~Mukunoki, A.~Podobas, M.~Wahib, and S.~Matsuoka, ``Matrix {Engines} for
  {High} {Performance} {Computing}: {A} {Paragon} of {Performance} or
  {Grasping} at {Straws}?'' in \emph{2021 {IEEE} {International} {Parallel} and
  {Distributed} {Processing} {Symposium}, {IPDPS} 2021, {Portland}, {Oregon},
  {USA}, {May} 17-21, 2021}, Portland, Oregon, USA, May 2021, p.~10.

\bibitem{hoefler_scientific_2015}
T.~Hoefler and R.~Belli, ``Scientific benchmarking of parallel computing
  systems: twelve ways to tell the masses when reporting performance results,''
  in \emph{Proceedings of the {International} {Conference} for {High}
  {Performance} {Computing}, {Networking}, {Storage} and {Analysis}}, ser. {SC}
  '15, Austin, TX, USA, Nov. 2015, pp. 73:1--73:12.

\bibitem{christina_blog_2020}
\BIBentryALTinterwordspacing
T.~Christina, ``Blog {Post}: {Significant} performance improvements in {GCC} 10
  through {Vectorization} and {In}-lining,'' May 2020. [Online]. Available:
  \url{https://community.arm.com/developer/tools-software/tools/b/tools-software-ides-blog/posts/gcc-10-better-and-faster}
\BIBentrySTDinterwordspacing

\bibitem{huber_case_2021}
\BIBentryALTinterwordspacing
J.~Huber, W.~Wei, G.~Georgakoudis, J.~Doerfert, and O.~R. Hernandez, ``A {Case}
  {Study} of {LLVM}-{Based} {Analysis} for {Optimizing} {SIMD} {Code}
  {Generation},'' Tech. Rep. arXiv:2106.14332v1, 2021. [Online]. Available:
  \url{https://arxiv.org/pdf/2106.14332.pdf}
\BIBentrySTDinterwordspacing

\bibitem{burford_ookami_2021}
\BIBentryALTinterwordspacing
A.~Burford, A.~C. Calder, D.~Carlson, B.~M. Chapman, F.~CoSKun, T.~Curtis,
  C.~Feldman, R.~J. Harrison, Y.~Kang, B.~Michalow-Icz, E.~Raut, E.~Siegmann,
  D.~G. Wood, R.~L. DeLeon, M.~Jones, N.~A. Simakov, J.~P. White, and
  D.~Oryspayev, ``Ookami: {Deployment} and {Initial} {Experiences},'' Tech.
  Rep. arXiv:2106.08987v1, 2021. [Online]. Available:
  \url{https://arxiv.org/pdf/2106.08987.pdf}
\BIBentrySTDinterwordspacing

\bibitem{michalowicz_comparing_2021}
B.~Michalowicz, E.~Raut, Y.~Kang, T.~Curtis, B.~M. Chapman, and D.~Oryspayev,
  ``Comparing the behavior of {OpenMP} {Implementations} with various
  {Applications} on two different {Fujitsu} {A64FX} platforms,'' in
  \emph{Practice and {Experience} in {Advanced} {Research} {Computing}}, ser.
  {PEARC} '21.\hskip 1em plus 0.5em minus 0.4em\relax New York, NY, USA:
  Association for Computing Machinery, 2021.

\bibitem{poenaru_evaluation_2021}
\BIBentryALTinterwordspacing
A.~Poenaru, ``An {Evaluation} of the {Fujitsu} {A64FX} for {HPC}
  {Applications},'' Presentation in AHUG ISC 21 Workshop, Jul. 2021. [Online].
  Available: \url{https://a-hug.org/isc-2021-event/}
\BIBentrySTDinterwordspacing

\bibitem{alappat_performance_2020}
C.~Alappat, J.~Laukemann, T.~Gruber, G.~Hager, G.~Wellein, N.~Meyer, and
  T.~Wettig, ``Performance {Modeling} of {Streaming} {Kernels} and {Sparse}
  {Matrix}-{Vector} {Multiplication} on {A64FX},'' in \emph{2020 {IEEE}/{ACM}
  {Performance} {Modeling}, {Benchmarking} and {Simulation} of {High}
  {Performance} {Computer} {Systems} ({PMBS})}, 2020, pp. 1--7.

\bibitem{sreepathi_e3sm_2021}
\BIBentryALTinterwordspacing
S.~Sreepathi, O.~Guba, and M.~Taylor, ``{E3SM} {Pathfinding} on {Fugaku},'' May
  2021. [Online]. Available: \url{https://e3sm.org/e3sm-pathfinding-on-fugaku/}
\BIBentrySTDinterwordspacing

\bibitem{jackson_investigating_2020}
A.~Jackson, M.~Weiland, N.~Brown, A.~Turner, and M.~Parsons, ``Investigating
  {Applications} on the {A64FX},'' in \emph{2020 {IEEE} {International}
  {Conference} on {Cluster} {Computing} ({CLUSTER})}.\hskip 1em plus 0.5em
  minus 0.4em\relax Los Alamitos, CA, USA: IEEE Computer Society, Sep. 2020,
  pp. 549--558.

\bibitem{odajima_preliminary_2020}
T.~Odajima, Y.~Kodama, M.~Tsuji, M.~Matsuda, Y.~Maruyama, and M.~Sato,
  ``Preliminary {Performance} {Evaluation} of the {Fujitsu} {A64FX} {Using}
  {HPC} {Applications},'' in \emph{2020 {IEEE} {International} {Conference} on
  {Cluster} {Computing} ({CLUSTER})}, 2020, pp. 523--530.

\end{thebibliography}

%
%
%

\end{document}